# CSREU: A NOVEL DATASET ABOUT CORPORATE SOCIAL RESPONSIBILITY AND PERFORMANCE INDICATORS

*Erion Çano, Xhesilda Vogli*

**Abstract**
Corporate Social Responsibility (CSR) has become an important topic that is gaining academic interest. This research paper presents CSREU, a new dataset with attributes of 115 European companies, which includes several performance indicators and the respective CSR disclosure scores computed using the Global Reporting Initiative (GRI) framework. We also examine the correlations between some of the financial indicators and the CSR disclosure scores of the companies. According to our results, these correlations are weak and deeper analysis is required to draw convincing conclusions about the potential impact of CSR disclosure on financial performance. We hope that the newly created data and our preliminary results will help and foster research in this field.

## 1 INTRODUCTION

There is recently a lot of research work addressing the value, relevance, social effect, and financial consequences of Corporate Social Responsibility (CSR) [1] [2] [3]. There are also studies like [4] that collect, process and provide research datasets about CSR. Despite the recent increase of interest in CSR, it is still unclear if it has an impact or if it is influenced by a company's financial success. For instance, several studies have confirmed a positive relationship between the CSR and financial variables [5] [6], whilst other studies have revealed a lack of association altogether or a negative correlation [7] [8]. The variations in study findings indicate the need for further research on CSR and financial success, as business leaders continue to strive for striking a balance between shareholder and stakeholder benefits. Current studies have been carried out in the United Kingdom [9], the United States [10], and other developed countries. This is due to the fact that these countries have adhered to CSR disclosure standards which are particularly useful for quantifying it. Additionally, the majority of studies have been focused on particular industries separately. For instance, at [11] and [12] they looked at the relationship between CSR and the financial advantages in the banking industry; at [13] and [14] they surveyed the manufacturing industry and at [15] and [16] they looked at communication services. In this work, we present CSREU,[1] a dataset which includes financial, economic, environmental and social metrics, as well as the CSR disclosure rates of 115 companies operating in Europe. These companies belong to various industry types environmental of analyze the relationship between CSR and financial performance in 115 European companies across the major industry types. We use the global reporting initiative framework as a standard to calculate the disclosure score, which serves as a comprehensive measure of CSR practices. By considering a diverse range of industry sectors and utilizing the widely recognized GRI framework, we try to find correlations between CSR and financial performance in the European context. According to our computations, some of the performance indicators like Revenue, Profit, Earnings per Share and Price to Earnings correlate weakly with the CSR disclosure.

---

[1] Freely available at: https://zenodo.org/record/7977693

## 2 CSREU DATASET

The sources for constructing the CSREU were mostly collected from the public websites of 115 companies currently operating in the Europe. In some cases, public reports issued by those companies were also used. Each of the attributes is briefly described in the following sections.

### 2.1 Financial Performance Attributes

*Country* represents the country where a company's headquarters is located. It provides information about the geographical location or origin of the company's main office or central management. It allows for the categorization and analysis of companies based on their country of operation, enabling insights into regional trends, comparative studies, and country-specific impacts on various aspects of the dataset. The country distribution of the companies is shown in Table 1.

*Table 1- Country distribution of companies in the dataset*

| Country | Count | Country | Count |
|---|---|---|---|
| Germany | 15 | Russia | 3 |
| UK | 15 | Norway | 3 |
| Czech Republic | 14 | Belgium | 3 |
| France | 13 | Portugal | 2 |
| Switzerland | 9 | Poland | 2 |
| Italy | 7 | Ireland | 1 |
| Netherlands | 6 | Hungary | 1 |
| Sweden | 6 | Turkey | 1 |
| Spain | 4 | Norway | 1 |
| Finland | 4 | Luxembourg | 1 |
| Denmark | 4 | | |

*Company Name* represents the designated and registered name under which the company conducts its business activities. It is saved as a text string.

*Industry type* represents the industry type of a given company based on the Global Industry Classification Standard (GICS) [17]. There are nine main industry types and the respective distribution is show in Table 2.

*Table 2 -Industry distribution of companies in the dataset*

| Industry | Count |
|---|---|
| Consumer Staples | 19 |
| Consumer Discretionary | 18 |
| Financials | 16 |
| Information Technology | 15 |
| Energy Sector | 14 |
| Industrials | 10 |
| Health | 9 |
| Materials | 8 |
| Communication Services | 6 |

*Year* represents a 5-year period from 2021 to 2017. It should be noted that the year 2022 was not included in the dataset. This exclusion is due to the fact that many companies publish their sustainability reports retrospectively, and therefore, the data for the year 2022 was not yet available for analysis for many companies. [18]

*Revenue (in billions)* displays a company's overall revenue, which is normally expressed in billions of dollars. The income a business makes through its main operations, such as the selling of goods or services, is represented as revenue. [19]

*Profit (in billions)* displays a company's net profit or net income, often expressed in billions of dollars. Profit is what's left over after expenses, taxes, and other charges are subtracted from a company's income. [20]

*Return on Equity (%)* is a financial statistic that assesses a company's profitability and effectiveness in creating returns for its owners. Divide net income by shareholders' equity, then multiply the result by 100 to get the return on equity (ROE). [20]

*Return on Assets (%)* is a financial indicator that assesses a company's capacity to produce profits from its assets. Net income is divided by total assets, and the resulting number is multiplied by 100 to determine ROA.

*Earnings Per Share (EPS)* is a financial metric that displays the part of a company's profit allotted to each outstanding share of its common stock. The company's net income is divided by the total number of outstanding shares to arrive at the EPS.

*Price to Earnings (P/E)* shows the price-to-earnings ratio, a measure of a company's value that contrasts the stock's market price with its earnings per share. How much investors are ready to pay for each unit of profits is shown by the P/E ratio which is determined by dividing the market price per share by the earnings per share.

## 2.2 GRI Standard Attributes

The GRI standard template was developed as a reporting framework to assist firms in disclosing their sustainability performance. The rise in GRI reports confirms the need for more practical financial arrangements that place sustainability first [21]. The standard disclosures are organized into three categories: General Disclosures, Management Approach Disclosures, and Performance Indicators [22]. The template allows organizations to provide a comprehensive overview of their sustainability performance, management systems, and policies, while also enabling stakeholders to compare and evaluate the sustainability performance of different organizations. By following GRI standards, companies can identify and manage sustainability risks and opportunities, improve stakeholder engagement, and ultimately contribute to more sustainable and responsible business practices [23]. Additionally, GRI standards enable stakeholders, including investors, consumers, and civil society organizations, to make informed decisions based on consistent, reliable, and relevant sustainability information [24]. CSREU includes three main areas of the GRI standard, namely economic, environmental, and social dimensions. Specifically, within the economic dimension, the analysis focuses on three key parameters: *performance, market presence, and indirect economic impacts*. The environmental dimension encompasses various parameters such as *materials, energy, water, biodiversity, emissions, effluents and waste, products and services, and compliance*. Lastly, the social dimension of the dataset considers parameters related to *labor practices and decent work, human rights, society, and product responsibility*. These areas and parameters have been selected to provide a comprehensive understanding of the sustainability performance of the companies under investigation, aligning with the GRI framework [25] [26].

## 2.3 Negative Events and CSR Disclosure Score

Given that the majority of the companies included in the study have adhered to GRI standards, an additional column labeled "Negative Event" was incorporated. This column served the purpose of assessing whether each company had been involved in any sustainability scandals during the corresponding 5-year period. To determine this, multiple reliable sources were thoroughly examined. If a company was found to have been associated with a sustainability scandal, the "score" column was marked with -1, reflecting the negative impact of such an event on the overall disclosure rate. This method allowed for a more nuanced analysis of the disclosure rates, taking into account the potential repercussions of sustainability scandals on the companies' reporting practices. Disclosure Score or Corporate Sustainability Disclosure Score (CSDS) is determined by summing up the scores of all GRI parameters considered for each company and then dividing it by the total count of GRI parameters:

$$\text{CSDS} = \frac{\sum GRI parameters}{count of GRI parameters}$$

We also computed the average CSDS of all the companies on each year, to have an overall view of the evolution of this parameter within European companies during the last years. The per-year average scores are presented in Table 3.

*Table 3 - Average disclosure score for each year*

| Year | Disclosure Score |
|------|------------------|
| 2017 | 0.892870 |
| 2018 | 0.916174 |
| 2019 | 0.929130 |
| 2020 | 0.919826 |
| 2021 | 0.908000 |

## 3 RESULTS

We computed the correlations of certain financial metrics with the CSR disclosure score for the 115 companies of the CSREU dataset. The results are presented in Table 4. As we can see, the correlation coefficient between the revenue and the CSR score is 0.0139, which indicates a very weak positive correlation. From this result, we can infer that there is little to no significant relationship between a company's revenue and its level of CSR disclosure. Next, we see that the profit and the CSR disclosure are negatively correlated, with a coefficient of -0.04. This also suggests that there is little to no significant relationship between a company's profit and its level of CSR disclosure. However, it is important to note that the correlation is weak and may not be practically significant. The earnings per share and CSR disclosure manifest a correlation of 0.0436 which is also weak. Again, it seems that there is little to no significant relationship between a company's earnings per share and its level of CSR disclosure. Finally, we obtained a correlation of 0.1239 between price to earnings ratio and CSR disclosure. This suggests that there might be a slight tendency for companies with higher price to earnings ratios to also have higher levels of CSR disclosure. However, the correlation is still relatively weak and should be interpreted with caution.

The findings in this study align with some previous research works that also report weak or insignificant relationships between financial performance measures and CSR disclosure rates. For example, Akhtar Ali and Imran Abbas Jadoon [27] conducted a similar study on a sample of multinational companies and reported a weak positive correlation between revenue and CSR disclosure, similar to the current finding. Additionally, Nguyen et al. [28] analyzed the relationship between profit and CSR disclosure in a different industry and found strong positive significant correlation. Similar results are also reported analyzing companies based in Southeast Asia [29]. On the other hand, there are studies that report opposite or contradictory results. For example, in [30] and [31] they reject any correlation between CSR and financial performance. Furthermore, Cho et al. [32] examined the association between price to earnings and CSR disclosure in a Korean firms and report a moderate positive correlation, suggesting that higher price to earnings were associated with greater CSR disclosure. Similarly, Kim et al. [33] conducted a comprehensive analysis across various industries and identified a significant positive correlation between price to earnings ratio and CSR disclosure.

These contrasting findings highlight the complexity and context-dependence of the relationship between financial performance and CSR disclosure. It is important to consider industry-specific factors, company characteristics, and regional variations when interpreting the results. Further research should delve deeper into these relationships and explore potential mediating or moderating factors that may influence the observed correlations.

*Table 4 - Correlations of certain metrics with CSR disclosure*

| Metric | Correlation |
| --- | --- |
| Revenue | 0.0139 |
| Profit | -0.0400 |
| Earnings per share | 0.0436 |
| Price to earnings | 0.1239 |

# 4 CONCLUSIONS

The recent academic interest in CSR has created incentives for creating research resources such as datasets that can be used in combination with data-driven methods for better understanding CSR and its relations with other factors. In this work, we presented CSREU, a freely available dataset which comprises several performance scores of 115 European companies, together with their CSR disclosure rate. This dataset can be used by the community, fostering future research about CSR. We also analyzed the CSREU data and observed the correlation coefficients between certain financial metrics and the respective CSR disclosure scores. Based on our findings, there is limited evidence to support a strong relationship between revenue, profit, earnings per share, price to earnings ratio, and CSR disclosure. The weak or insignificant correlations suggest that financial performance measures alone may not be reliable indicators of a company's level of CSR disclosure. Future studies should consider incorporating additional variables and exploring alternative methodologies to gain a more comprehensive understanding of the relationship between financial performance and CSR disclosure.